\documentclass[aps,prc,twocolumn,amsmath,amssymb,superscriptaddress,showpacs,showkeys,preprintnumbers]{revtex4-2}
\usepackage{dcolumn}
\usepackage{graphicx,color}
\usepackage{multirow}
\usepackage{textcomp}
\usepackage{physics}
\usepackage{dsfont}
\usepackage{comment}
\usepackage{relsize}

\usepackage{tikz}	
\begin{document}
	\newcommand {\nc} {\newcommand}
	\nc {\beq} {\begin{eqnarray}}
	\nc {\eeq} {\nonumber \end{eqnarray}}
	\nc {\eeqn}[1] {\label {#1} \end{eqnarray}}
\nc {\eol} {\nonumber \\}
\nc {\eoln}[1] {\label {#1} \\}
\nc {\ve} [1] {\mbox{\boldmath $#1$}}
\nc {\ves} [1] {\mbox{\boldmath ${\scriptstyle #1}$}}
\nc {\mrm} [1] {\mathrm{#1}}
\nc {\half} {\mbox{$\frac{1}{2}$}}
\nc {\thal} {\mbox{$\frac{3}{2}$}}
\nc {\fial} {\mbox{$\frac{5}{2}$}}
\nc {\la} {\mbox{$\langle$}}
\nc {\ra} {\mbox{$\rangle$}}
\nc {\etal} {\emph{et al.}}
\nc {\eq} [1] {(\ref{#1})}
\nc {\Eq} [1] {Eq.~(\ref{#1})}
\nc {\Refc} [2] {Refs.~\cite[#1]{#2}}
\nc {\Sec} [1] {Sec.~\ref{#1}}
\nc {\chap} [1] {Chapter~\ref{#1}}
\nc {\anx} [1] {Appendix~\ref{#1}}
\nc {\tbl} [1] {Table~\ref{#1}}
\nc {\Fig} [1] {Fig.~\ref{#1}}
\nc {\ex} [1] {$^{#1}$}
\nc {\Sch} {Schr\"odinger }
\nc {\flim} [2] {\mathop{\longrightarrow}\limits_{{#1}\rightarrow{#2}}}
\nc {\textdegr}{$^{\circ}$}
\nc {\inred} [1]{\textcolor{red}{#1}}
\nc {\inblue} [1]{\textcolor{blue}{#1}}
\nc {\IR} [1]{\textcolor{red}{#1}}
\nc {\IB} [1]{\textcolor{blue}{#1}}
\nc{\pderiv}[2]{\cfrac{\partial #1}{\partial #2}}
\nc{\deriv}[2]{\cfrac{d#1}{d#2}}
\nc{\gsim}{\raisebox{-0.13cm}{~\shortstack{$>$ \\[-0.07cm]
      $\sim$}}~}
\nc {\bit} {\begin{itemize}}
	\nc {\eit} {\end{itemize}}

\title{Systematic study of the validity of the eikonal model including uncertainties}
\author{D.~Shiu}
\email{shiudani@msu.edu}
\affiliation{Facility for Rare Isotope Beams, Michigan State University, East Lansing, Michigan 48824, USA}
\affiliation{Department of Physics and Astronomy, Michigan State University, East Lansing, Michigan 48824, USA}
\author{C.~Hebborn}
\email{hebborn@frib.msu.edu}
\affiliation{Université Paris-Saclay, CNRS/IN2P3, IJCLab, 91405 Orsay, France}
\affiliation{Facility for Rare Isotope Beams, Michigan State University, East Lansing, Michigan 48824, USA}
\affiliation{Department of Physics and Astronomy, Michigan State University, East Lansing, Michigan 48824, USA}
\author{F.~M.~Nunes}
\email{nunes@frib.msu.edu}
\affiliation{Facility for Rare Isotope Beams, Michigan State University, East Lansing, Michigan 48824, USA}
\affiliation{Department of Physics and Astronomy, Michigan State University, East Lansing, Michigan 48824, USA}

\date{\today}

\begin{abstract}
\begin{description}
\item[Background] Nuclear reactions at intermediate beam energies are often interpreted using the eikonal model. In the analysis of complex reaction probes, where few-body reaction methods are needed, the eikonal method may be used as an efficient way for describing the fragment-target reaction process. 
\item[Purpose] In this work, we perform a systematic study to test the validity of the eikonal approximation for nucleon-nucleus reactions. We also quantify uncertainties due to the nucleon optical potential on reaction observables.
\item[Method] We inspect the validity of the eikonal model and its semiclassical correction by comparing it to  exact solutions (obtained from solving the optical model equation with a finite differences method) for a wide range of reactions. We also study the effect of relativistic corrections, both kinematic and dynamic, by effectively incorporating the relativistic effects at intermediate energies. 
The uncertainties from a Bayesian global optical potential (KDUQ) are propagated to the observables of interest.
\item[Results] Our study includes neutron and proton reactions on $^{27}$Al, $^{40}$Ca, $^{90}$Zr and $^{208}$Pb, for a wide range of energies $E_{lab}=0-400$ MeV. We calculate neutron total cross sections (elastic and reactions) as well as proton absorption cross sections as a function of beam energy, using the eikonal model, the eikonal model with a semiclassical correction, and the exact solution. We also compute angular distributions for the  methods above. 
\item[Conclusions] Our results show that for the proton absorption cross section, the eikonal model can be used down to around $60$ MeV and the semiclassical correction extends its use to $30$ MeV. However, the validity of the eikonal model for the neutron total cross section only goes down to $\approx120$ MeV, a range extended to $\approx 50$ MeV when using the semiclassical correction. We find the semi-classical correction to the eikonal model to be less effective in describing the angular distributions.
The $1\sigma$ uncertainty intervals on the observables we studied is less than $5$\% for most of the energies considered, but increases rapidly for higher energies, namely energies outside the range of KDUQ ($E_{lab}>200$ MeV).
\end{description}
\end{abstract}

\maketitle

\section{Introduction}
\label{sec:intro}

Nuclear reaction experiments and theory are crucial to  study the  properties of exotic nuclei located away from stability. Being short-lived, most exotic nuclei cannot be assembled into a target, but are instead synthesized and smashed onto a stable target. Combining accurate reaction measurements and theory to interpret the experimental data, one can extract  properties of the colliding nuclei  (see recent reviews in Ref.~\cite{Hammachereview,annual2020,KayReview}). These reactions are typically described within few-body models, composed of clusters of nucleons,  and take as input effective interactions between its constituents, the so-called optical potentials (see recent review~\cite{op2023}). To make accurate reaction predictions, it is crucial to quantify the  uncertainties associated with the few-body reaction models and   the optical potentials used as inputs. 

Solving the  few-body dynamics becomes increasingly complex when more degrees of freedom are retained in the model. Hence, it is common to use approximations to describe reactions involving three or more clusters~\cite{ThompsonNunesBook,Satchler,BertulaniBook}.   At high energy, i.e., above 50$A$~MeV, the eikonal approximation is a tool of choice as it reduces drastically the computational time, it is easily generalizable to collisions involving multiple clusters and it provides a simple semi-classical interpretation of the collision:  the projectile is seen as propagating along straight-line trajectories along which it accumulates a phase through its interaction with the target~\cite{G59,HT03,BC12}. Expectedly, the eikonal model becomes inaccurate at low energies, where  straight-line trajectories do not make sense~\cite{PhysRevC.96.054607,PhysRevC.98.044610}.  Because of its  advantages, several works have studied corrections to the eikonal model to better account for the deflection of the projectile by the target,  hence extending its  range of validity. These corrections were derived from perturbative expansions~\cite{Wal73,Wal71,WalPhD},  from semiclassical arguments~\cite{PhysRevC.46.1026,LenziVitturiZardi95,PhysRevC.90.034617,PhysRevC.56.1511,PhysRevC.96.054607,PhysRevC.98.044610,PhysRevC.109.014621,HebbornPhD} and from correspondence to the exact solution~\cite{BrookePhD,Wal73,BAT99}. 
Although the eikonal model and its corrections   were previously analyzed for specific applications (e.g. see Refs.~\cite{PhysRevC.96.054607,PhysRevC.98.044610}), there is no comprehensive study of their accuracy for  reactions on various targets over a broad range of energies.

Moreover,  reaction model predictions  carry parametric uncertainties associated with the optical potentials taken as input~\cite{lovell2018,king2018,lovell2021,catacora2023,KDUQ,pruitt2024,SmithUQchex,KOUQ22,prl2023,prl2023err}. 
Recently, studies have shown that the magnitude of the uncertainty in the reaction due to optical-model uncertainties depends on the reaction probe. An example is the work analyzing one-nucleon transfer and knockout reactions \cite{prl2023,prl2023err}). In that work,  the same optical model uncertainties produce credible intervals with different widths in the knockout cross sections and in the transfer cross sections. We underline the fact that the model used for knockout relied on the eikonal solution for the two-body scattering amplitudes whereas the model used for transfer did not (the adiabatic  wave approximation (ADWA)~\cite{JT74,NLAT} takes exact two-body amplitudes as input). One possible justification for the differences seen is that the optical-model uncertainties propagate differently through the two-body eikonal than through the exact solver.

This work aims at addressing these two features: it clarifies the range of  validity of the eikonal approximation and its correction, and it compares how optical uncertainties propagate in the eikonal model and the exact finite-differences method. 
Across all the reactions here considered, we use the recently-developed uncertainty quantified KDUQ global optical model~\cite{KDUQ}. This global nucleon-nucleus optical potential was calibrated using a Bayesian analysis on hundreds of scattering datasets on a variety of targets, at a wide range of beam energies and a number of observables.  In this work, we compare cross sections obtained with the eikonal model (and the corrected eikonal model)  with the exact method. We consider  absorption and elastic integrated cross sections, as well as elastic angular distributions for protons and neutrons on  $^{27}$Al,  $^{40}$Ca, $^{90}$Zr and  $^{208}$Pb at energies  up to 400~MeV. 

This paper is organized as follow. In Section~\ref{sec:TheoStats}, we briefly describe the two-body exact model along with the eikonal method and its correction, as well as the statistics concepts used in this work. The cross sections obtained with all three methods are presented in \ref{sec:results}, with additional results shown in Appendix. Also in this section, we discuss the validity of the eikonal approximation, its correction and the propagation of the potential parameters uncertainties. Finally, Section~\ref{sec:conclusions} contains the conclusions of this work.

\section{Theoretical and statistical frameworks}
\label{sec:TheoStats}

We consider the scattering of a nucleon projectile $P$ of mass $m_P$ off a nucleus target $T$ of mass $m_T$.  We model this reaction as a two-body system, in which the target $T$ is assumed structureless and spinless. Within this framework, the interaction of the nucleon with the target is described by an optical potential $U$ containing both a real and an imaginary component as well as the Coulomb field $U=V_N+iW+V_C$.

In the exact solver, the  wavefunction $\Psi$ is obtained from the solution of the \Sch equation
\begin{equation}
    \left[ T_{\ve r} +  U(\ve r)\right] \Psi(\ve r)= E\,\Psi(\ve r),\label{EqSch}
\end{equation}
where $T_{\ve r}$ is the kinetic operator, $\ve r \equiv(\ve b, z)$ is the relative $P$-$T$ coordinate, $\ve b$ and $z$ are the transverse and longitudinal $P$-$T$ coordinates.  We solve this equation considering the initial condition that the projectile is impinging onto the target with a momentum $\ve k=k\ve z$, that we choose along the $z$-axis.

In this work, we use the uncertainty global  nucleon-nucleus  optical potential KDUQ~\cite{KDUQ}.  The potential parameters were calibrated using a Bayesian approach onto various scattering observables at energies between 0 and 200 MeV. 
At intermediate energies ($\gsim 100$ MeV), relativistic effects can become important and need to be included~\cite{KDUQ,KD03,Ogata_Bertulani_2010,Moschini_Capel_2019}. We follow the same approach as in the KDUQ calibration~\cite{KDUQ}, which is described in Ref.~\cite{Ingemarsson} (equations 17 and 20). First, the center-of-mass energies, angles, and relative velocity are calculated using relativistic kinematics. Second, the optical potential is scaled by a factor derived from a comparison between the Klein-Gordon equation and the \Sch equation.

The exact solution to this \Sch equation and the exact cross sections are  obtained with the partial-wave expansion method. In this method, one solves the partial-wave radial \Sch equation and the $S$-matrix for each partial wave  are deduced by a matching condition with the known asymptotic form. All observables are computed from these $S$-matrices. The exact calculations shown in Sec. \ref{sec:results} are obtained  with the  code {\sc frescox}~\cite{Fresco}.

\subsection{Eikonal model}
The eikonal model  reflects the fact that, at high enough energy, the projectile is only slightly deflected by the target.  In this model, the wavefunction is factorized into the initial plane wave and a new wavefunction $\hat{\Psi}^{\rm eik}$
\begin{equation}
    \Psi(\ve r) \approx e^{ikz} \hat{\Psi}^{\rm eik}(\ve r).
\end{equation}
Using this factorization and  neglecting the  second derivative of $\hat{\Psi}^{\rm eik}$, the \Sch equation~\eqref{EqSch} is simplied into
\begin{equation}
    i\hbar v \frac{\partial}{\partial z} \hat{\Psi}^{\rm eik}(\ve b,z)=U(b,z) \hat{\Psi}^{\rm eik}(\ve b,z),
\end{equation}
where $v=\hbar k/\mu $ is the initial velocity.
This simplified \Sch  equation can be solved analytically
\begin{equation}
    \Psi^{\rm eik}=e^{ikz}e^{-\frac{i}{\hbar v}\int_{-\infty}^z U(b,z')dz'}.
\end{equation}
The eikonal solutions can be interpreted semiclassically as  the projectile following a straight-line trajectory, at constant impact parameter $b$, and accumulating a phase through its interaction  $U$ with the target.

The observables are then  computed from the eikonal $S$-matrix
\begin{eqnarray}
    S^{\rm eik}(b)=e^{-\frac{i}{\hbar v}\int_{-\infty}^{+\infty} U(b,z)dz}.
\end{eqnarray}
Compared to the exact $S$-matrix, which are obtained by a matching condition with the known asymptotic form, the eikonal $S$-matrices  depend on the integration of the potential at all distances. Typically, the short-range parts of phenomenological optical potentials are not well constrained~\cite{Igo,SmithUQchex}. It is hence unclear if the uncertainties of the optical potential propagate similarly in the eikonal model and in the exact method.   We address  this question in the next section by comparing the uncertainties on various reaction observables computed with the eikonal approximation  to the ones obtained with the exact method.

\subsection{Semi-classical correction to the eikonal model}

The eikonal model becomes inaccurate at low energies, when the deflection of the projectile by the target is important and  straight-line trajectories do not make sense. To extend the validity of the eikonal approximation, one can use a semiclassical correction, which shifts the impact parameter $b$ to the classical distance of closest approach~$b'$
\begin{equation}
    S^{\rm eik}(b) \rightarrow S^{\rm eik}(b').
\end{equation}
Previous studies show that this correction efficiently corrects for the deflection due to the Coulomb interaction~\cite{PhysRevC.90.034617} and is often used in eikonal analyses~\cite{BertulaniBook,SuzukiBook}. 

One can generalize this correction to the nuclear interaction by using  the distance of closest approach derived from both the Coulomb and the real part of the nuclear potential $V_N+V_C$. It was shown that this correction extends  the validity of the eikonal model down to slightly lower energies~\cite{PhysRevC.46.1026,LenziVitturiZardi95,ProceedingsHebborn}. Nevertheless, these previous works focus on specific applications.   Here, a systematic study of this real semiclassical correction is performed, over a large range of energies, targets and observables.

 A further generalization of this  correction to include a complex distance of approach (obtained from including both the real and imaginary part of the optical potential) was later proposed~\cite{PhysRevC.56.1511,PhysRevC.96.054607,PhysRevC.98.044610,PhysRevC.109.014621}. However, because the interpretation of a complex distance is unclear, and because it leads to an overestimation of the absorption in three-body collision~\cite{PhysRevC.98.044610,HebbornPhD}, we do not consider this generalization in our current study.

 \subsection{Statistical considerations}

 To evaluate the uncertainties associated with the optical potentials onto the reaction observables, we use the 416 samples of the KDUQ democratic posterior distribution~\cite{KDUQ}. From these 416 cross sections, we compute  the credible intervals as the smallest   interval that include $x\%$ of the cross section predicted by the 416 samples of KDUQ. In the eikonal calculations, we did not include the KDUQ spin-orbit terms, whose treatment is non trivial~\cite{SuzukiBook}. By running the exact calculations with and without spin-orbit terms, we verified that, for the observables here considered, the spin-orbit term can be neglected.

 In order to compare how uncertainties propagate in each model, we compare in the next section the size of the the half-width $\varepsilon_{68}$ of the $68\%$ credible interval  $[\sigma^{{68}}_{min},\sigma^{{68}}_{max}]$, defined as
 \begin{eqnarray}
     \varepsilon_{68}&=&\frac{\sigma^{{68}}_{max}-\sigma^{{68}}_{{avg}}} {\sigma_{{avg}}}\label{halfwidth_eq}\\
    \text{with}\quad  \sigma^{{68}}_{{avg}}&=&\frac{\sigma^{{68}}_{max}+\sigma^{{68}}_{min}}{2}.
 \end{eqnarray}

\section{Results} 
\label{sec:results}

This systematic study includes calculations for protons and neutrons on various stable targets, representing a wide range of masses: $^{27}$Al,  $^{40}$Ca, $^{90}$Zr and  $^{208}$Pb. For each case, we calculate the angle-integrated elastic and absorption cross sections as a function of beam energy up to $E_{lab}=400$ MeV. In addition to the cross sections, we compute angular distributions for elastic scattering. In this section, we only present results for $^{208}$Pb and discuss this case in detail. The results for the other targets are shown in Appendix A. 

All observables were checked for convergence. The eikonal calculations for the integrated cross sections were performed by integrating the trajectory along the beam axis between  $[-z_{max},z_{max}]$ with $z_{max}=10$~fm, with a grid for impact parameter   between $[0,b_{max}]$ with $b_{max}=15$~fm.
For the angular differential cross sections, we used $z_{max}=30$~fm and $b_{max}=60$~fm.
For the exact solver, we take partials waves up to $L=400$ and for each partial wave integrate the equation up to $R_{max}=60$~fm with a radial step size of $R_{step}=0.01$~fm.

\begin{figure*}[htb!]
    \centering
    
    \includegraphics[width=0.75\linewidth]
    {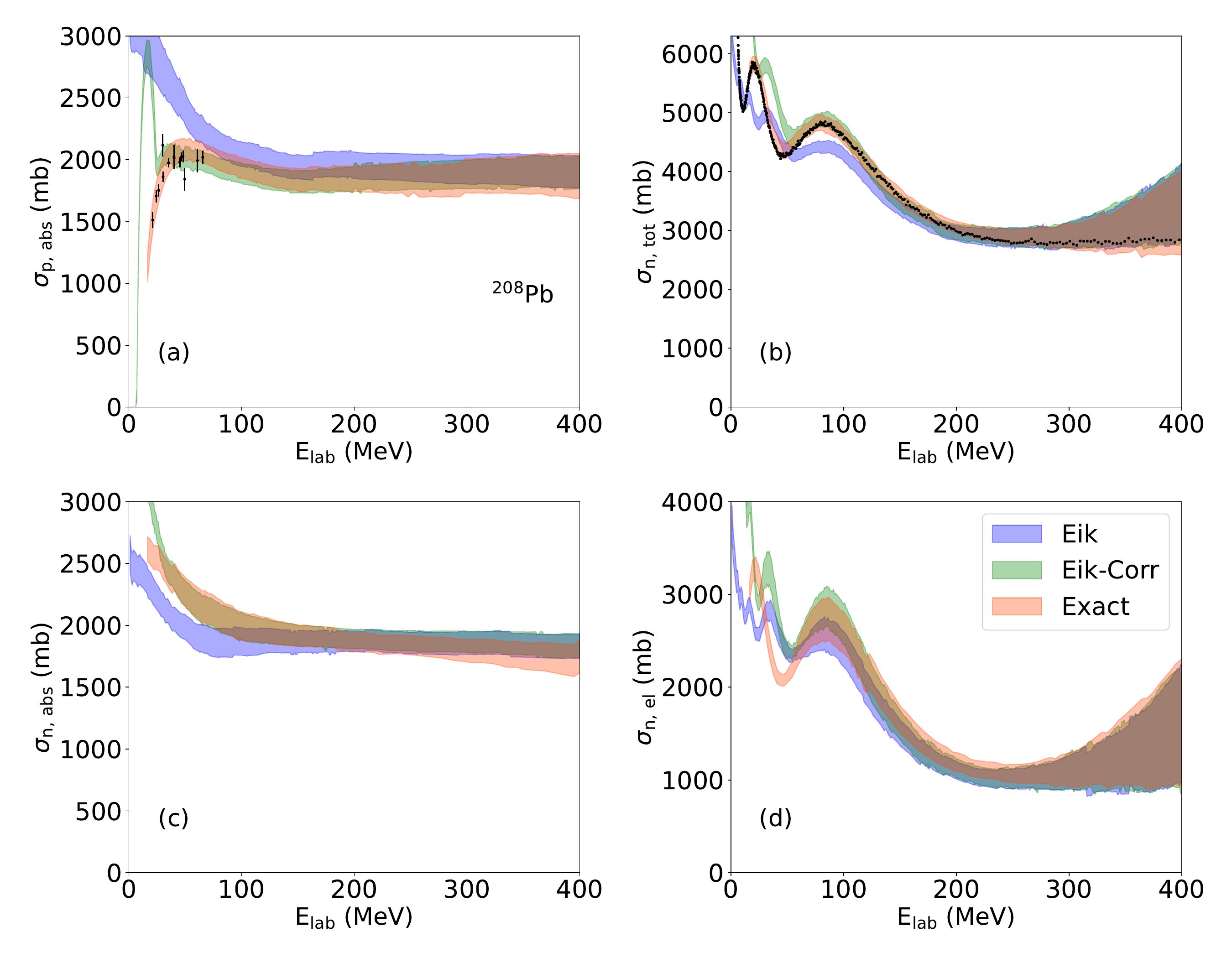}
    \caption{Cross sections as a function of beam energy for nucleon scattering on $^{208}$Pb: (a) proton absorption cross section; (b) neutron total cross section; (c) neutron absorption cross section; and (d) neutron elastic cross section. The 68\% credible intervals are shown for the Eikonal model (blue), the Eikonal model with semi-classical correction (green) and the exact solution (orange). Also plotted are the data from Refs.~\cite{PhysRevC.47.237,PhysRevC.12.1167,TURNER1964509,PhysRevC.4.1114,INGEMARSSON1999341}.}
    \label{fig:1}
\end{figure*}

We first analyze the $68$\% credible intervals for the angle-integrated cross sections as a function of the nucleon beam energy. Fig. \ref{fig:1}  includes, on the left, the absorption cross section for either protons (a) or neutrons (c) on  $^{208}$Pb. The panels on the right of Fig. \ref{fig:1} are  (b) the neutron total cross section and (d) the neutron elastic cross section. Calculations for both the standard eikonal model (blue) and the version including the semi-classical correction (green) are shown. Also displayed are the exact results (orange), along with the data. Since the KDUQ optical potential was fitted on this data, it is no surprise that the exact results align well with the data.

\begin{table}[t!]
\centering
\renewcommand{\arraystretch}{1.2}
\setlength{\tabcolsep}{6pt}
\begin{tabular}{|ll||c|c|}
\hline
& &{\textbf{$E_{min}^{\rm eik}$}} (MeV) & {\textbf{$E_{min}^{\rm eik-corr}$}}  (MeV) \\
\hline
\multirow{5}{*}{$\sigma_{\rm p, abs}$}&\textsuperscript{208}Pb & 60  & 25    \\
&\textsuperscript{90}Zr & 35 & 30    \\
&\textsuperscript{40}Ca & 25  & 25   \\
&\textsuperscript{27}Al & 25  & 25  \\
\hline \hline
 \multirow{5}{*}{$\sigma_{\rm n, tot}$}&\textsuperscript{208}Pb  & 110  &  50  \\
&\textsuperscript{90}Zr & 160  & 40  \\
&\textsuperscript{40}Ca & 150 & 50    \\
&\textsuperscript{27}Al & 120   & 40   \\\hline
\end{tabular}
\caption{Comparison of energies in MeV from which the exact 68\%  confidence interval starts to overlap with the ones obtained in  the eikonal model (eik) and its semiclassical correction (eik-corr) using  relativistic kinematics and dynamics. }\label{table}
\end{table}

\subsection{Validity of the eikonal approximation}

First, we compare the Eikonal model (blue band in Fig.~\ref{fig:1}) with the exact solution (orange band in Fig. \ref{fig:1}). Expectedly, the two models agree at high energies and the eikonal model breaks down towards lower energy. However, we find that the energy below which the eikonal model breaks down $E_{min}$  varies with the observable and the target. In Table~\ref{table}, we compiled $E_{min}^{\rm eik}$, corresponding to the beam energy below which the deviation between the eikonal predictions and the exact cross sections are noticeable to the eye. For proton cross sections, the energy range changes by factors of 3 from the lightest target ($E_{min}^{\rm eik}=25$~MeV) to the heaviest target ($E_{min}^{\rm eik}=60$~MeV).   This suggests that the deflection of the trajectory by the Coulomb interaction, neglected in the eikonal model,  should be corrected for.

For neutrons the eikonal model is only valid up to around $E_{min}^{eik} \approx 120$~MeV, a value much higher than often assumed. 
The strong difference between $E_{min}^{eik}$ for neutrons and protons hints at the importance of the Coulomb component in the scattering amplitude; this component is treated exactly in the eikonal model.
We also note that the eikonal model underestimates the total neutron cross section, a result already noted in Ref.~\cite{PhysRevC.108.044609}.

\subsection{Validity of the semi-classical approximation}

When introducing the semi-classical correction, the eikonal predictions improve significantly (green bands). The values of the beam energy below which the corrected eikonal deviates from the exact calculations is represented by $E_{min}^{\rm eik-corr}$ in Table~\ref{table}. The large dependence of the range of validity with target mass disappears when including the semi-classical correction, particularly for $\sigma_{p,abs}$. 
As for the pure eikonal, the validity of the corrected eikonal method extends to much lower energy for $\sigma_{p,abs}$ than for $\sigma_{n,tot}$. Indeed, at low energy (below 50 MeV), the correction on $\sigma_{n,tot}$ overshoots, predicting a much larger cross section for $E_{lab}<50$ MeV.

\subsection{Relativistic corrections}

It is important to note that all calculations shown in Fig. \ref{fig:1} include relativistic corrections. We have performed a detailed study comparing the relativistic effects themselves, both the kinematic correction and the dynamic correction (equations 17 and 20 of Ref.~\cite{Ingemarsson}).
Namely, we have compared the results obtained with the eikonal model, with and without relativistic corrections, to the exact solutions, with and without these corrections. A summary of this systematic study is included  in Appendix~\ref{SecRelativitistic}. We found that  these corrections become significant above $E_{lab}=100$ MeV, for the observables discussed in Fig. \ref{fig:1} and  Table~\ref{table}. 

\subsection{Elastic angular distributions}

\begin{figure*}[htb!]
    \centering
    \includegraphics[width=\linewidth]{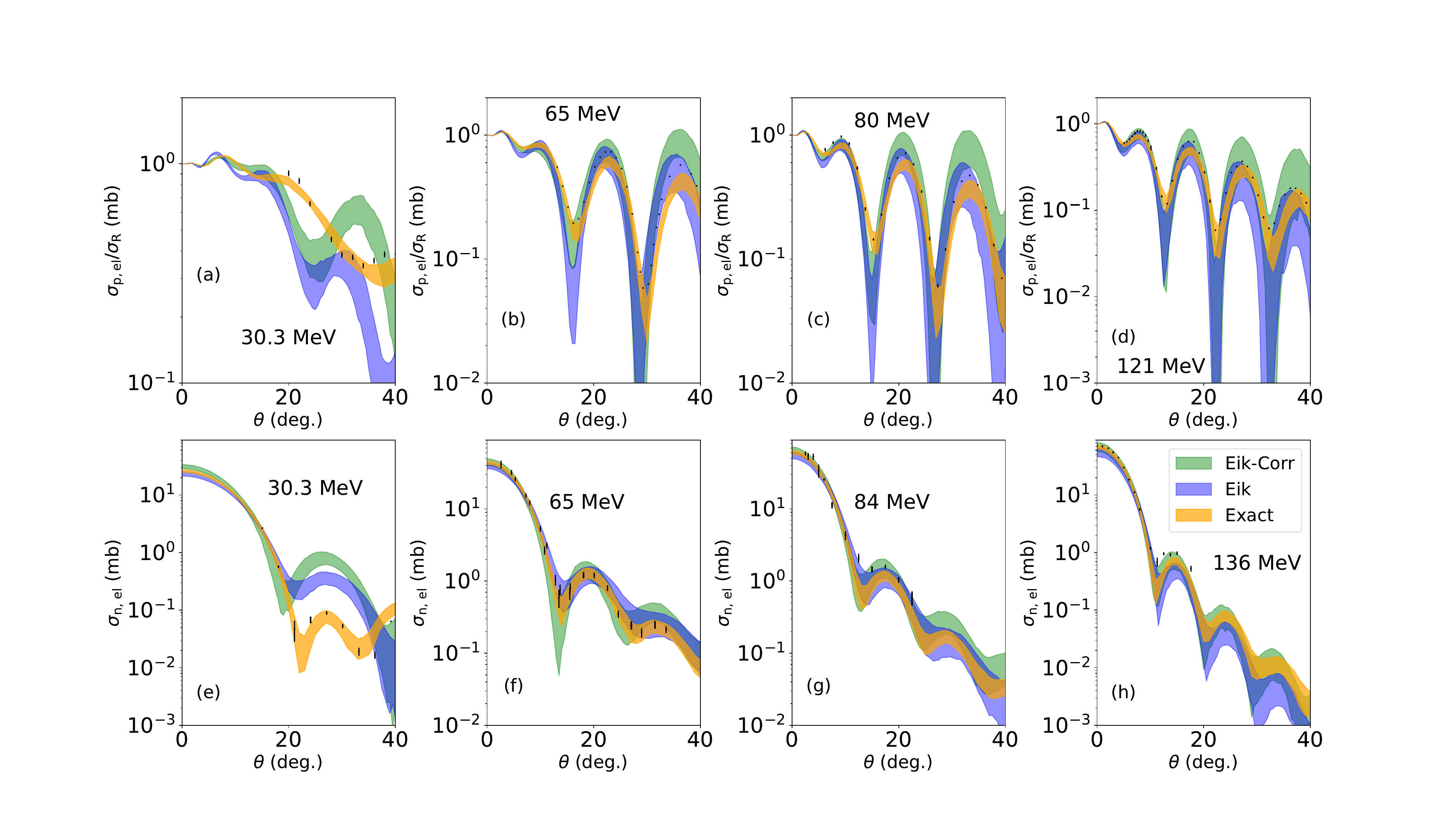}

    \caption{Cross section angular distributions as a function of scattering angle for nucleon elastic scattering on $^{208}$Pb: (a) proton scattering at 30.3 MeV, (b) proton scattering at 65 MeV; (c) proton scattering at 80 MeV; (d) proton scattering at 136 MeV; (e) neutron scattering at 30.3 MeV, (f) neutron scattering at 65 MeV; (g) neutron scattering at 84 MeV; and (h) neutron scattering at 136 MeV.
The 68\% credible intervals are shown for the eikonal model (blue), the eikonal model with semi-classical correction (green) and the exact solution (orange). Also plotted are the data from Refs.~\cite{DataprotonElScat,PhysRevC.26.944,PhysRevC.23.1023,DataneutronElScat,Baba01082002,VanZyl01111956,PhysRev.77.597}.}
    \label{fig:2}
\end{figure*}

We now turn to the elastic angular distributions shown in Fig. \ref{fig:2}. The top (bottom) row corresponds to proton (neutron) elastic scattering. The different panels include beams energies for which there is  data, starting at $E_{lab}=30$~MeV up to  $E_{lab}=136$~MeV. Again the exact calculations (orange) were used to constrain KDUQ and therefore are able to reproduce the experimental diffraction patterns (black circles). At the highest energy here considered, the eikonal model matches the angular distributions obtained with the exact model. It is only at the lowest energy plotted here that the eikonal approximation fails to describe the diffraction pattern. For such cases, unfortunately, the semi-classical correction offers little to no improvement.

\subsection{Propagation of uncertainties}

\begin{figure*}[htb!]
    \centering
    \includegraphics[width=0.8\linewidth]{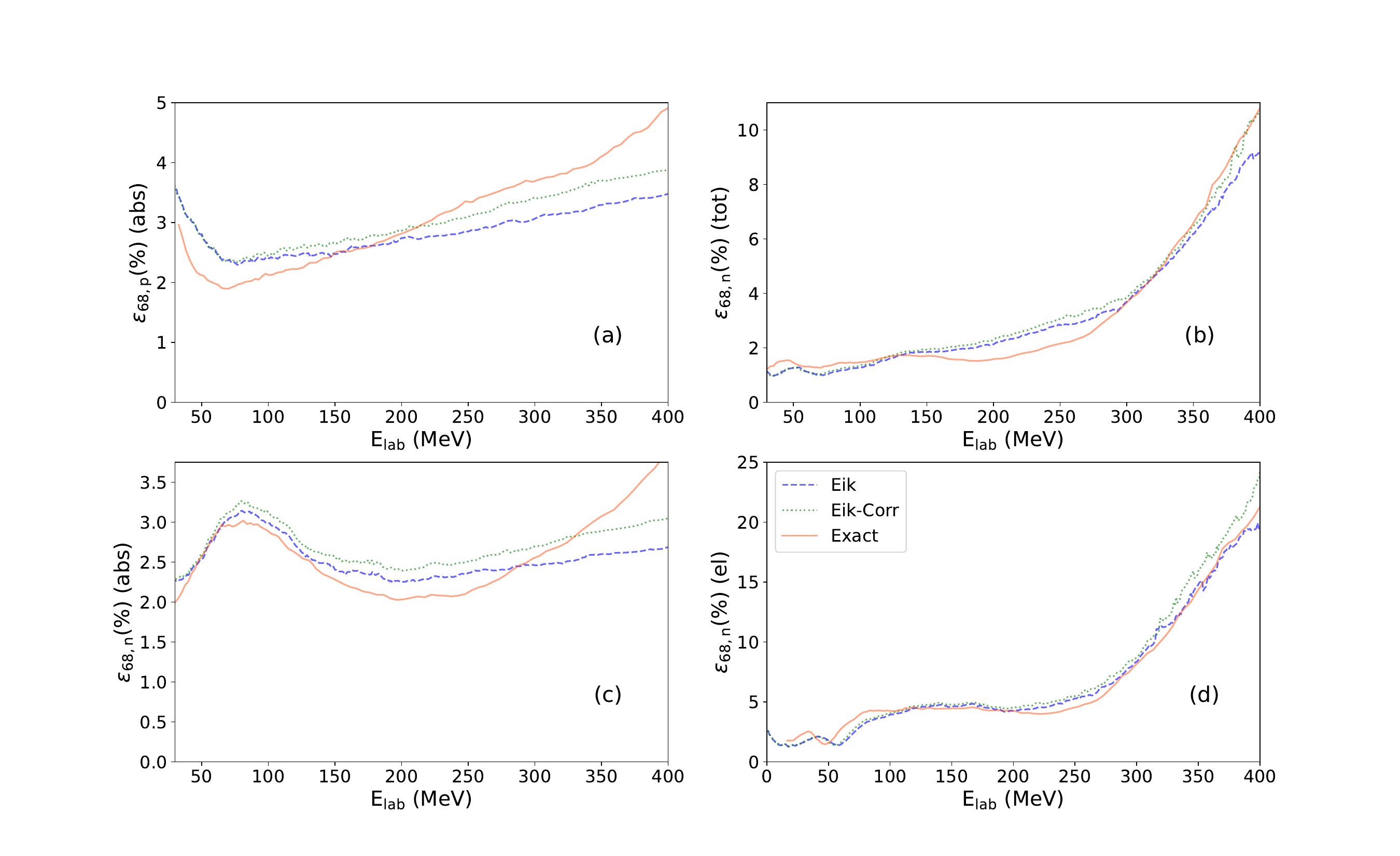}
    \caption{Percent $1\sigma$ error on the cross sections as a function of beam energy for each panel shown in Fig.\ref{fig:1}: (a) proton absorption; (b) neutron total; (c) neutron absorption and (d) neutron elastic. Included are the results with the eikonal model (blue), the corrected eikonal (green) and the exact (orange).}
    \label{fig:3}
\end{figure*}

We now inspect the effect of optical model uncertainties when propagated through the eikonal models versus the exact model. Given the two methods are foundationally different (one involving an impact parameter integration while the other involved a partial wave decomposition and matching of the solutions at large distances) it is unclear whether the two models would propagate optical model uncertainties in a similar manner.
Shown in Fig. \ref{fig:3} are the relative widths of the 68\% credible intervals presented in Fig. \ref{fig:1}. We focus first at energies below $200$ MeV: for all considered cases, the uncertainties on the cross section are small ($<5$\%). Much of this data has been used to constrain KDUQ and therefore it is expected that the uncertainties in this region are small.

Next we discuss the case for which the uncertainties are largest, namely the neutron elastic cross sections for $E_{lab}>200$ MeV. The eikonal models predict similar uncertainties to the exact, and most importantly, the uncertainties increase rapidly  reaching $25$\% for the neutron elastic cross section at $E_{lab}=400$~MeV. This sudden increase starts  at $E_{lab}=200$~MeV, which  is precisely the maximum energy of validity of the KDUQ parameterization.   When investigating the energy dependence of KDUQ, we find that for higher energies, the energy dependence of the real depth of  KDUQ (see Eq.(A9) of \cite{KDUQ}) is poorly constrained (specifically the coefficient of $\Delta E^4=(E-\epsilon_f)^4$ is very uncertain; $-4.30^{+25.60}_{-20.30}\times 10^{-9}$). This has an impact in the neutron elastic cross sections. To reduce these uncertainties, the  KDUQ parametrization should be extended to higher energies.

As the neutron elastic cross section dominates the neutron total cross section,  the uncertainties in the neutron total cross section shown in  panel (b) have a similar energy behavior as those shown in panel (d), with all models predicting identical uncertainties. Concerning the neutron and proton absorption cross sections, the error remains small across the whole energy range considered here ($<5$\%), as these cross sections depend on the imaginary part of the optical potential and are not affected by the real term in KDUQ discussed before. Although there are some differences in the estimated uncertainties when propagated through the three reaction models for the absorption cross section, these differences are not significant.

\section{Conclusions}
\label{sec:conclusions}

We perform a systematic study of the validity of the eikonal model and its semiclassical correction for a large number of reactions.
Results for nucleon-nucleus reactions on stable targets with masses $A=27-208$ and for energies up to $E_{lab}=400$ MeV are calculated, both in the eikonal model and using an exact solver. All our results use the KDUQ global optical potential for the nucleon-nucleus effective interaction. This global potential  has uncertainties quantified  through a rigorous Bayesian analysis. We propagate these uncertainties to all considered reaction observables.

We first study the integrated cross sections as a function of beam energy. For proton-target reactions we find that, for $^{208}$Pb, the validity of the eikonal model ranges down to $60$ MeV, with the semiclassical correction extending this region of validity down to $30$ MeV. For the neutron-target reactions, the validity of the eikonal model is much reduced (only down to $150$~MeV). In this case, the semiclassical correction extends it down to $50$~MeV. Note that the eikonal model has also been applied to reactions around $70A$~MeV~\cite{GadeTostevin,Moschini_Capel_2019,PhysRevC.104.024616,BC12,PhysRevC.109.014621}. In such applications, it is imperative to use the semiclassical correction for a valid interpretation of the measurements.

We also analyzed the angular distributions for elastic scattering. For proton (neutron) distributions,  the eikonal model is not able to capture the diffraction pattern accurately when compared to the exact solutions below $70$ MeV ($40$ MeV). We find that, for the angular distributions, the semi-classical correction is not helpful in bringing the eikonal predictions closer to the exact solutions. Finally, we also looked at the need for relativistic corrections. In the cases considered, we found that relativistic corrections become significant for $E_{lab}>100$~MeV. 

Concerning the magnitude of the uncertainties due to the optical potentials on the relevant reaction observables, we find that uncertainties estimated with the eikonal model are similar to those obtained with the exact solver. For most of the energy range considered, the uncertainties on the cross sections are small ($<5$ \%) (a result consistent with a recent study of transfer reactions~\cite{FrontiersHebbornNunes}) but they grow rapidly for the higher energies we included in this study. We note that the corpus of data constraining the KDUQ parameterization does not include reactions with $E>200$ MeV. This is the region where our predicted uncertainties become considerable $\approx 25$\%. One must also note that all reactions considered in this study involve stable targets. KDUQ was only constrained by stable-target data. One should expect larger  uncertainties in extrapolating the quantified uncertainties to rare isotopes~\cite{op2023}. 

When analyzing complex reactions, often we need nucleon-target reaction S-matrices for a wide range of energies. As alluded to before, these may dip below the region of validity of the eikonal model. Some codes switch over abruptly from the fully quantal partial-wave decomposition method into the eikonal one, at an ad-hoc energy. Our studies provides concrete values for the energy where this transition should occur. Perhaps more interestingly, it opens the prospect of using Bayesian model mixing to smoothly describe this transition while keeping track of the uncertainties.

\begin{acknowledgements} The authors thank Kyle Beyer for the discussion on the energy dependence of the real depth of KDUQ.
  The work of F.M.N. was in part supported by the U.S. Department of Energy grant DE-SC0021422 and National Science Foundation CSSI program under award No. OAC-2004601 (BAND Collaboration).
\end{acknowledgements}

\bibliographystyle{apsrev}
\bibliography{Biblio}
\appendix 

\section{Cross sections on $^{27}$Al, $^{40}$Ca and $^{90}$Zr targets}
\begin{figure*}[htb!]
    \centering
    \includegraphics[width=0.75\linewidth]{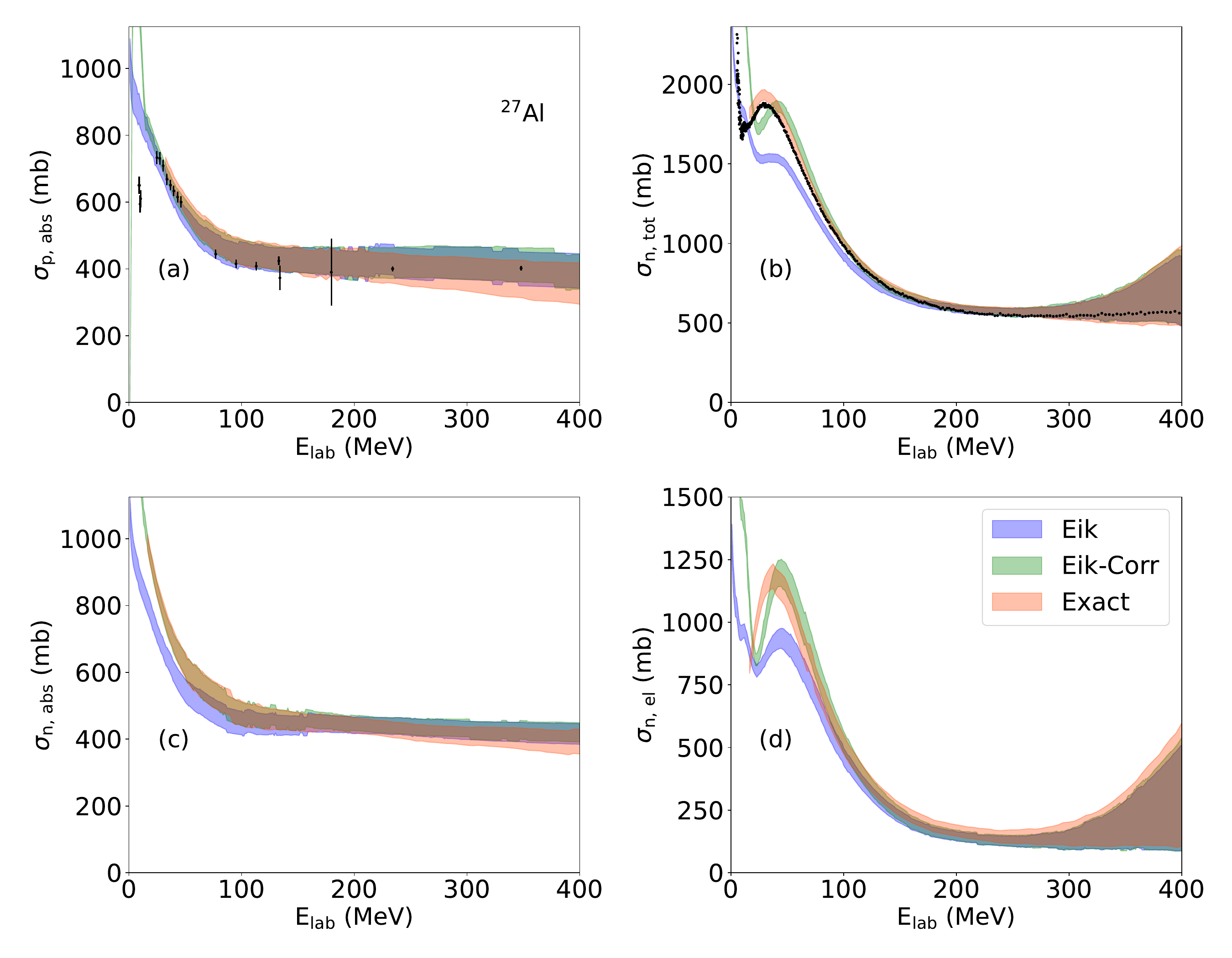}
    \caption{Same as Fig. \ref{fig:2} for a $^{27}$Al target. Also plotted are the data from Refs.~\cite{PhysRevC.47.237,BEARPARK1965206,PhysRev.140.B575,PhysRevC.10.2237,MAKINO1964145,GOODING1959241,PhysRevC.4.1114,PhysRev.117.1334,GOLOSKIE1962474,JMCassels_1954,exforp27Al,RENBERG197281}.}
    \label{fig27Al}
\end{figure*}

\begin{figure*}[htb!]
    \centering
    \includegraphics[width=0.75\linewidth]{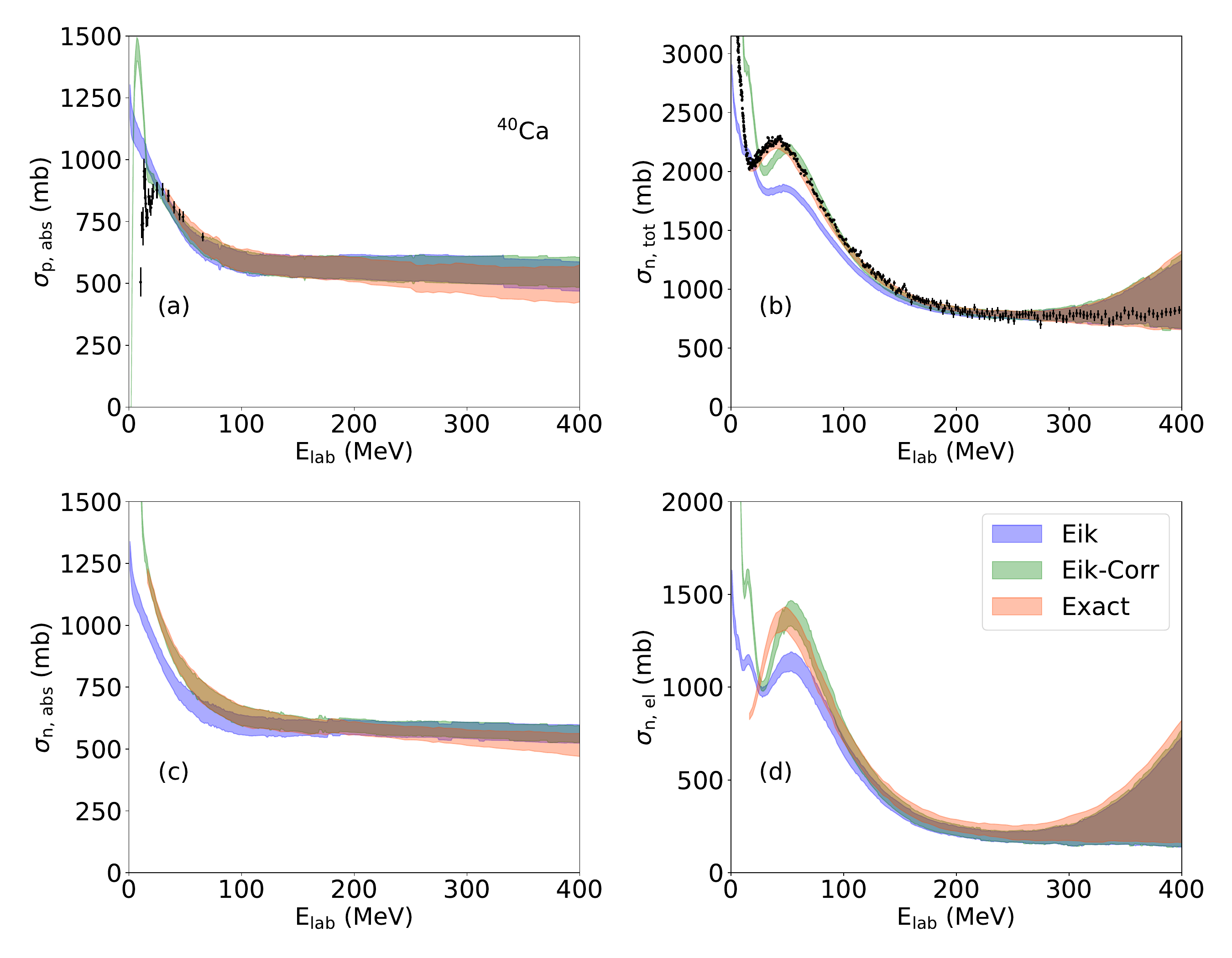}
    \caption{Same as Fig. \ref{fig:2} for a $^{40}$Ca target. Also plotted are the data from Refs.~\cite{PhysRevC.47.237,PhysRevC.2.488,PhysRevC.12.1167,TURNER1964509,INGEMARSSON1999341}.}
    \label{fig40Ca}
\end{figure*}

\begin{figure*}[htb!]
    \centering
    \includegraphics[width=0.75\linewidth]{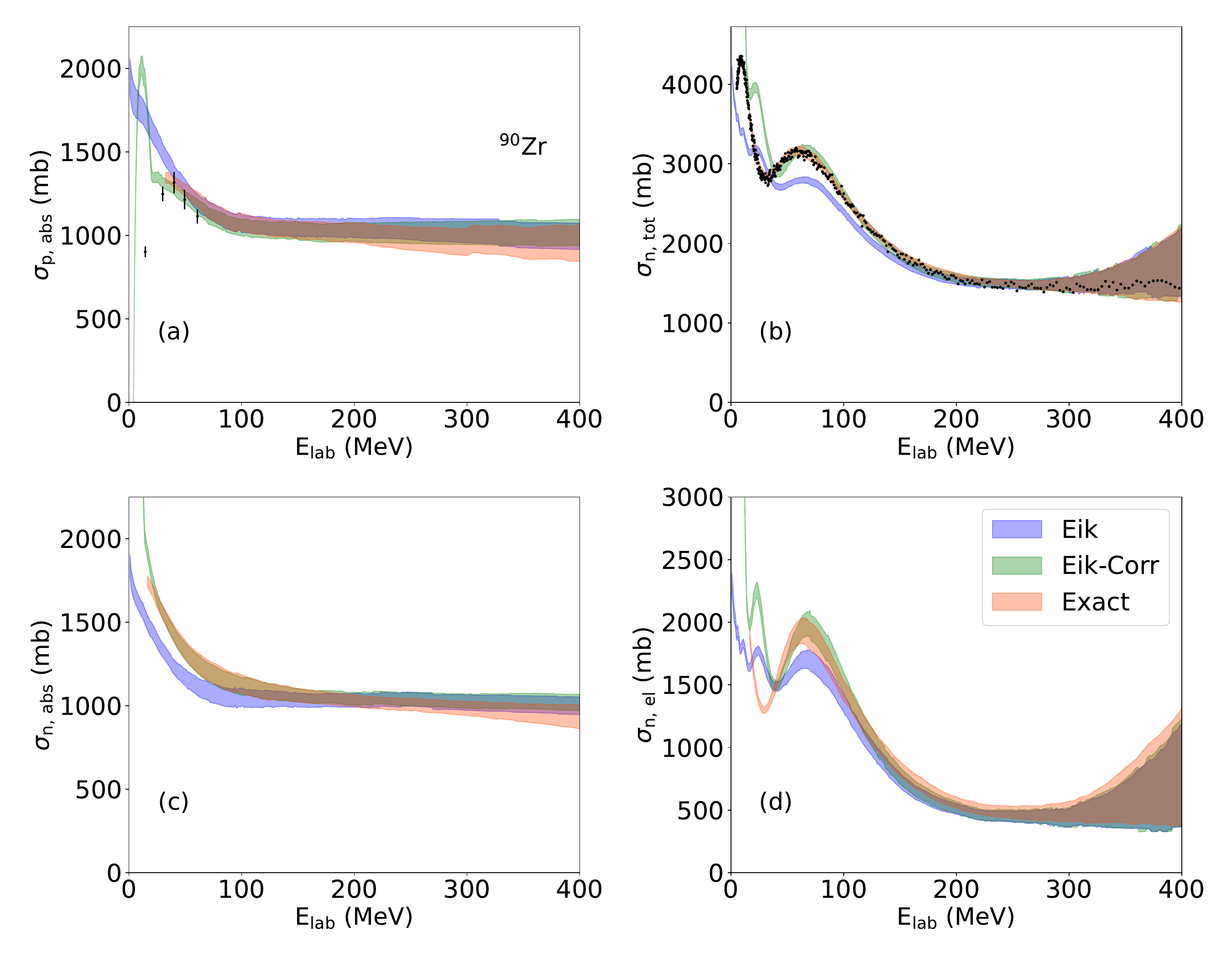}
    \caption{Same as Fig. \ref{fig:2} for a $^{90}$Zr target. Also plotted are the data from Refs.~\cite{PhysRevC.47.237,PhysRevC.4.1114,PhysRev.157.1001}.}
    \label{fig90Zr}
\end{figure*}

For completeness, this appendix presents the remaining results included in our study. All figures are identical to Fig.~\ref{fig:2} in the main text, but now for the following targets: $^{27}$Al (Fig.~\ref{fig27Al}), $^{40}$Ca (Fig.\ref{fig40Ca}) and $^{90}$Zr (Fig.\ref{fig90Zr}). The conclusions from the joint analysis of all these figures is included in the main text.

\section{Importance of relativitic kinematics and dynamics} \label{SecRelativitistic}

 \begin{figure*}[htb!]
    \centering
    \includegraphics[width=0.75\linewidth]{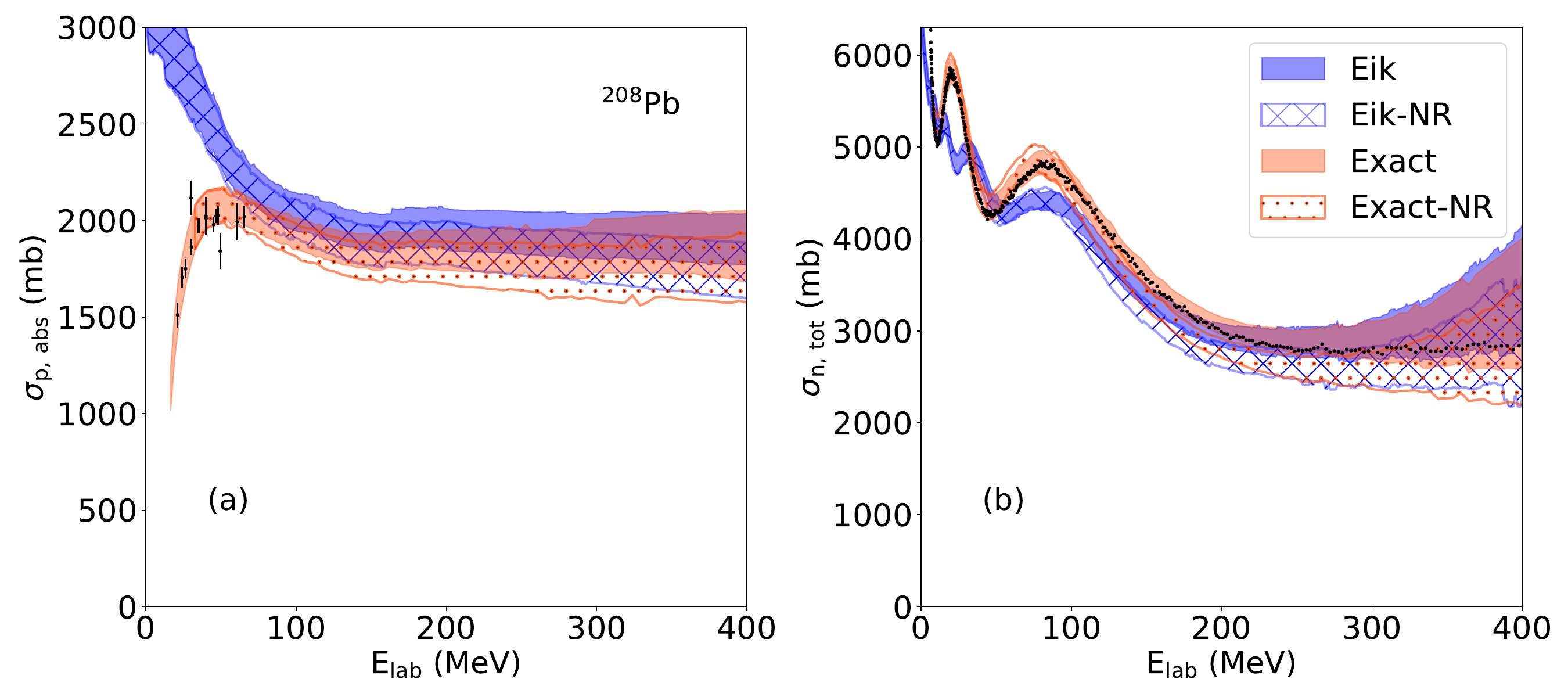}
    \caption{Study of the relativistic correction for proton absorption (left) and neutron total (right) cross sections on a $^{208}$Pb target. We compare the exact (Exact, orange bands) and eikonal (Eik, blue bands) cross sections obtained  using relativistic~\cite{Ingemarsson} (filled bands) and non-relativistic (NR, dashed and dotted bands)  kinematics and dynamics (see Sec.~\ref{sec:TheoStats} for details). }
    \label{fig:4}
\end{figure*}
In this appendix, we study the importance of the relativistic kinematics and dynamics corrections in proton absorption and neutron total cross sections. We show in Fig. \ref{fig:4} the cross sections  on a $^{208}$Pb target obtained in the exact method (orange bands) and the eikonal model (blue bands), comparing the non-relativistic results with the ones obtained with both corrections. As expected, the corrections are negligible at low energies, and become significant at $\gsim 100$ $A$ MeV for both models. Interestingly, together these corrections  act similarly on both models for both proton absorption and neutron total cross sections: it leads to an enhancement of both  cross sections at high energies, which is of similar magnitude in both eikonal and exact models, and is more important  for neutron-induced reactions than for proton-induced ones. At 400 MeV, the relativistic correction leads to an increase of $\sim 6-8\%$  in the proton absorption cross section and $\sim 20\% $ for the neutron total cross sections.

Analyzing these corrections in more detail, we found that the relativistic kinematics and dynamics corrections act in opposite ways: the relativistic kinematic correction leads to a decrease of the cross sections while the relativistic dynamic one enhances their magnitude at high energies. The magnitude of the  relativistic dynamic correction being larger, the overall effect is an increase of the cross sections at high energies.
We performed the same analysis for reactions on other targets and obtained similar  qualitative conclusions. Because relativistic corrections are significant for results at beam energies above 100$A$~MeV,  the results presented in Sec.~\ref{sec:results} are obtained including both relativistic corrections.

\end{document}